\documentclass[aps,amsmath,amssymb,twocolumn,floatfix,pra,footinbib,superscriptaddress,longbibliography]{revtex4-2}

\usepackage[svgnames]{xcolor}
 \usepackage[
            colorlinks=true,        
            allcolors = black,  
            citecolor=DarkBlue, 
            linkcolor=DarkBlue,
            urlcolor=DarkBlue
        ]{hyperref}            
\usepackage{amsmath, amsfonts, amssymb, amsthm, bbm}
\usepackage{dsfont}
\usepackage{enumitem}
\usepackage{bm} 
\usepackage{soul}
\usepackage{array}
\usepackage{mathtools}
\usepackage{graphicx}% Include figure files
\usepackage{epstopdf} %converting to PDF
\usepackage{epsfig}
\usepackage{dcolumn}% Align table columns on decimal point
\usepackage{bm}% bold math
\usepackage{times}
\usepackage{physics}
\usepackage{listings}
\usepackage{multirow}
\usepackage{colortbl}
\usepackage{float}
\usepackage{autobreak}
\usepackage{amsmath}
\usepackage{algorithm}
\usepackage[noend]{algpseudocode}
\usepackage{color}
\input{Qcircuit}

\usepackage{mleftright} %to get rid of the extra space around parentheses with \left and \right: write \mleft, \mright instead
\usepackage{units}

\newcommand{\be}{\begin{equation}}
\newcommand{\ee}{\end{equation}}
\newcommand{\bea}{\begin{eqnarray}}
\newcommand{\eea}{\end{eqnarray}}

\newcommand{\figref}[1]{\mbox{Fig.~\ref{#1}}}

\newcommand{\secref}[1]{\mbox{Section~\ref{#1}}}

\newcommand{\appref}[1]{\mbox{Appendix~\ref{#1}}}
\renewcommand{\eqref}[1]{\mbox{Eq.~(\ref{#1})}}

\newcommand{\figpanel}[2]{Fig.~\hyperref[#1]{\ref*{#1}(#2)}}
\newcommand{\figpanels}[3]{Fig.~\hyperref[#1]{\ref*{#1}(#2)-(#3)}}
\newcommand{\figpanelNoPrefix}[2]{\hyperref[#1]{\ref*{#1}(#2)}}
\begin{document}

\title{Gaussian conversion protocol for heralded generation of qunaught states}

\author{Yu Zheng}
\email{zhyu@chalmers.se}
\affiliation{Department of Microtechnology and Nanoscience, Chalmers University of Technology, 412 96 Gothenburg, Sweden}

\author{Alessandro Ferraro}
\affiliation{Centre for Theoretical Atomic, Molecular and Optical Physics, Queen's University Belfast, Belfast BT7 1NN, United Kingdom}
\affiliation{Dipartimento di Fisica ``Aldo Pontremoli,'' Universit\`{a} degli Studi di Milano, I-20133 Milano, Italy}

\author{Anton Frisk Kockum}
%\email{anton.frisk.kockum@chalmers.se}
\affiliation{Department of Microtechnology and Nanoscience, Chalmers University of Technology, 412 96 Gothenburg, Sweden}

\author{Giulia Ferrini}
\email{ferrini@chalmers.se}
\affiliation{Department of Microtechnology and Nanoscience, Chalmers University of Technology, 412 96 Gothenburg, Sweden}

\date{\today}

\begin{abstract}
%grid states
%Gottesman--Kitaev--Preskill (GKP) states, and more generally states with a grid phase-space structure, are prominent bosonic states allowing for  correcting errors when encoding qubits into oscillators and for enhanced sensing. These states are also resources for quantum computation: when supplemented to the set of Gaussian operations, they unlock the possibility of performing universal quantum computation. However, the preparation of such states in photonic systems remains a significant theoretical and experimental challenge. In this paper, we introduce a probabilistic Gaussian conversion protocol that allows for the heralded conversion of binomial states onto grid states. The existence of such a protocol implies that binomial states are resource states for universal quantum computation in combination with Gaussian  operations. \commale{the abstract above mainly discusses motivations, I would modify it to something like:}

In the field of fault-tolerant quantum computing, continuous-variable systems can be utilized to protect quantum information from noise through the use of bosonic codes. These codes map qubit-type quantum information onto the larger bosonic Hilbert space, and can be divided into two main categories: translational-symmetric codes, such as Gottesman--Kitaev--Preskill (GKP) codes, and rotational-symmetric codes, including cat and binomial codes.  The relationship between these families of codes has not yet been fully understood. We present an iterative protocol for converting between two instances of these codes --- GKP qunaught states and four-fold-symmetric binomial states corresponding to a zero-logical encoded qubit --- using only Gaussian operations. This conversion demonstrates the potential for universality of binomial states for all-Gaussian quantum computation and provides a new method for the heraladed preparation of GKP states. Through numerical simulation, we obtain GKP qunaught states with a fidelity of over \unit[98]{\%} and a probability of approximately \unit[3.14]{\%}, after only two steps of our iterative protocol, though higher fidelities can be achieved with additional iterations at the cost of lower success probabilities.
\end{abstract}

\maketitle

%-----------------------------------------------------------------------------

\section{Introduction}
\label{sec:level1}
In the context of fault-tolerant quantum computing, among the physical platforms able to host quantum information, continuous-variable systems offer a valid alternative to finite-dimensional ones~\cite{Braunstein:05, Weedbrook:12}. In fact, by exploiting the infinite-dimensional Hilbert space associated to each mode of a bosonic field, it is possible to protect, in a hardware-efficient manner, the quantum information carriers from the detrimental effects of noise. In particular, to achieve error-correction using continuous-variable systems, one resorts to bosonic codes, where qubit-type quantum information is encoded redundantly in the larger bosonic Hilbert space~\cite{chuang_bosonic_1997, gottesman_encoding_2001, ralph_quantum_2003, michael2016new, grimsmo_quantum_2020}.

A variety of bosonic codes have been introduced so far~\cite{joshi2021quantum, cai2021bosonic, albert2022bosonic}. At the level of single-mode encoding, two main families have been considered, endowed with distinct features related to their underlying symmetries in the quantum phase space: translational- and rotational-symmetric codes. The celebrated codes introduced by Gottesman, Kitaev, and Preskill (GKP codes)~\cite{gottesman_encoding_2001} belong to the first class; they display various attractive features, including the ability to correct any error and to allow for universal quantum computation with Gaussian operations alone, as well as excellent performance for sensing~\cite{duivenvoorden2017single} and quantum communication~\cite{noh2018quantum}. On the other hand, cat~\cite{cochrane1999macroscopically, mirrahimi2014dynamically} and binomial~\cite{michael2016new} codes possess rotational symmetry~\cite{grimsmo_quantum_2020}, which is specifically resilient to phase-insensitive losses --- one of the main source of noise in bosonic platforms, in particular in optics --- and they have been instrumental to achieve the first demonstration of quantum error correction beyond the break-even point~\cite{ofek2016extending}. The possible relations between these two families of codes have not been uncovered in full. In this work, we focus on binomial and GKP states, proving that the latter can be heralded from the former via protocols composed of Gaussian operations alone.   

Specifically, GKP states can be introduced as simultaneous eigenstates of two commuting displacement operators in position and momentum. The spacing between the peaks can be chosen such that a two-dimensional code space is available to encode a qubit~\cite{grimsmo2021quantum}. Like GKP states, a grid state~\cite{weigand_generating_2018}, or qunaught state~\cite{walshe_continuous-variable_2020}, is the simultaneous eigenstate of two displacement operators in position and momentum, but the spacing is chosen such that the eigenspace has dimension one, and therefore no quantum information is encoded (hence the name qunaught). However, these qunaught states can still be used for quantum error correction~\cite{walshe_continuous-variable_2020, larsen_fault_tolerant_2021}. Furthermore, a Bell pair of GKP states can be obtained by combining two qunaughts at a beam-splitter~\cite{walshe_continuous-variable_2020}, therefore entailing their universality for quantum computation with Gaussian operations alone. The qunaught states can also be used to prepare three-dimension cluster states, which support topological qubit error correction~\cite{noh_fault-tolerant_2020}. The protocol that we introduce here specifically targets qunaught states and uses, as inputs, binomial states encoding the logical qubit state $\ket{0}$. We highlight two features of our results. 

First, our scheme yields a Gaussian conversion protocol between two useful and experimentally motivated  non-Gaussian states, while only a handful examples of such protocols are available so far~\cite{vasconcelos_all-optical_2010, etesse_proposal_2014, weigand_generating_2018, zheng_gaussian_2021, PhysRevA.105.062446}. Indeed, from a theoretical viewpoint, our protocol demonstrates that binomial states corresponding to the encoded qubit $\ket{0}$ can provide a universal resource for all-Gaussian quantum computation, in the sense that they complement probabilistic Gaussian protocols~\cite{Ferraro:2005vh, adesso2014continuous, Albarelli:2018uu} to achieve universality. Previously, it was only known that binomial $\ket{0}$ states provided computational universality in combination with the preparation of other specific encoded binomial states (e.g., $\ket{+}$ and $\ket{H}$), along with complex non-Gaussian operations necessary to implement two-qubit gates~\cite{grimsmo_quantum_2020}. 
 
Second, our results provide a generation scheme for GKP states, which could become useful especially in the context of optics. In general, many protocols have been proposed with this purpose in a variety of architectures (see for example Refs.~\cite{terhal2020towards, fukui2022building} and references therein) and the first experimental demonstrations have been recently achieved in trapped ions~\cite{fluhmann_encoding_2019}  and superconducting microwave systems~\cite{campagne-ibarcq_quantum_2020, kudra_robust_2022,sivak_real-time_2022}. However in the context of photonics, where highly scalable architectures have been achieved~\cite{takeda2019toward, bourassa_blueprint_2021}, it is still a challenge to prepare such GKP states due to the requirement of highly non-linear operations. For instance, by exploiting strong interactions of properly shaped free electrons with light, optical GKP states with squeezing parameter above \unit[10]{dB} and fidelities above \unit[90]{\%}, with corresponding post-selection probability of \unit[10]{\%}, could be potentially generated~\cite{dahan_creation_2022}.
Using the non-linearity from cross-Kerr interaction, it has been shown that GKP states with \unit[10]{dB} squeezing could be generated with average fidelities of \unit[99.99]{\%} and \unit[99.9]{\%}, with corresponding success probabilities of \unit[2.7]{\%} and \unit[4.8]{\%}, respectively~\cite{fukui_generating_2022}.
%\commg{Is qunaught the same thing as the grid state in Etesse et al? Yu: Etesse et al refer GKP states as comb states. qunaught states are one-dimensianal grid/gkp states}

By substituting non-linear interactions with proper non-Gaussian input states, other schemes have been proposed. In particular, Vasconcelos {\it et al.}~\cite{vasconcelos_all-optical_2010} and Etesse {\it et al.}~\cite{etesse_proposal_2014} provided iterative schemes to ``breed'' --- i.e., add iteratively --- peaks of a qunaught state by using as input squeezed cat states in combination with linear optics and homodyne measurement with post-selection. %The key observation is that the combinations of the peaks in different position will return more peaks. 
%They showed that the breeding action can be understood in another way by computing the action of a symmetric beamsplitter (BS) on comb states~\cite{etesse_proposal_2014}.
However, the dependence on measurement result leads to the success probability decreasing rapidly as the number of iterations increase.
As an improvement over these protocols, Weigand {\it et al.}~\cite{weigand_generating_2018} showed that breeding can be related to phase estimation. After gradually projecting the input squeezed cat states onto an approximate eigenstate of the displacement operator as the approximate eigenvalue is learned through successive iterations of the protocol, a grid state is generated. This protocol becomes deterministic in the limit of a large number of iterations. 
However, it is an open question whether other input states than squeezed cat states allow for achieving qunaught states with Gaussian protocols.

The method that we introduce here is inspired by the one in  Ref.~\cite{weigand_generating_2018}. In contrast to the latter, our protocol converts input binomial states (instead of squeezed cat states) into a qunaught state. Our protocol projects the input binomial states onto the eigenstates of the displacement operators in the directions of both of the quadratures of the field, $\hat{q}$ and $\hat{p}$, yielding a qunaught state. This projection is achieved by successively measuring the $\hat q$ and $\hat p$ quadratures of the bosonic field, which progressively transforms the rotational symmetry into a translational one. We therefore refer to our protocol as the QP protocol.

The circuit that allows for implementing our protocol is composed of passive linear optics elements and homodyne detection, and all the non-linearity comes from the input binomial states. As said, such states have been generated experimentally in microwave cavities~\cite{axline2018demand, hu_quantum_2019, kudra_robust_2022, Eickbusch2022,ni_beating_2022}, and a proposal for all-optical experimental generation has also been given~\cite{nehra_all-optical_2021}, where in the latter case the nonlinearity necessary to generate such states is achieved by realizing a measurement of a nonlinear observable, combined with linear optical devices. 
Our protocol is able to generate qunaught states with \unit[4.95]{dB} squeezing with $>\unit[98]{\%}$ fidelity and success probability of $>\unit[3]{\%}$.

This article is organized as follows.
We first review some background concepts concerning binomial and qunaught states in \secref{sec:input-output-states}. Then we introduce our QP protocol in \secref{QP-breeding protocol}. In \secref{sec:result}, we show through numerical simulations that this protocol yields qunaught states and we analyze in detail its performance, characterizing how the fidelity of the generated states to the target qunaught state, along with its corresponding success probability, depend on various parameters.
%giving some intuition of how a protocol works without postselection. 
We close the paper with a summary and discussion in \secref{sec:conclusion}. 
The Appendixes provide complementary information on the effect of $\hat q$ and $\hat p$ measurements in the QP breeding protocol (Appendix~\ref{app:q-and-p-effect}), on the performance and features achieved after two iterations of the protocol depending on the measurement outcomes obtained (Appendixes~\ref{sec: all cases} and~\ref{app:best}), and finally on how the output fidelity depends on the order of the rotational symmetry, as well as on the truncation parameter of the input binomial states (Appendix~\ref{app:Numerical-N-K}).

%-----------------------------------------------------------------------------

\section{Input and target states: binomial and qunaught states}
\label{sec:input-output-states}

Here we briefly review the definitions of binomial and qunaught states, i.e., the input and target states, respectively, of the conversion protocol that we consider. 

%--------------------------------------------------------------

\subsection{Binomial states}

The 0-logical code-word of the $N$-fold binomial codes is defined as~\cite{michael2016new, albert_performance_2018}:
\begin{equation}
\label{eq:bino}
    \ket{0_{N}}=\sum^{\lfloor K/2 \rfloor}_{k=0}\sqrt{\frac{1}{2^{K-1}}{K \choose 2k}} \ket{2kN},
\end{equation}
where $\lfloor K/2 \rfloor$ is the floor function of $K/2$, with $K$ the truncation parameter, and $N$ is the order of the rotation symmetry, i.e., this state is invariant under a rotation by $\pi/N$.

\begin{figure}
\centering
\includegraphics[width=1\linewidth]{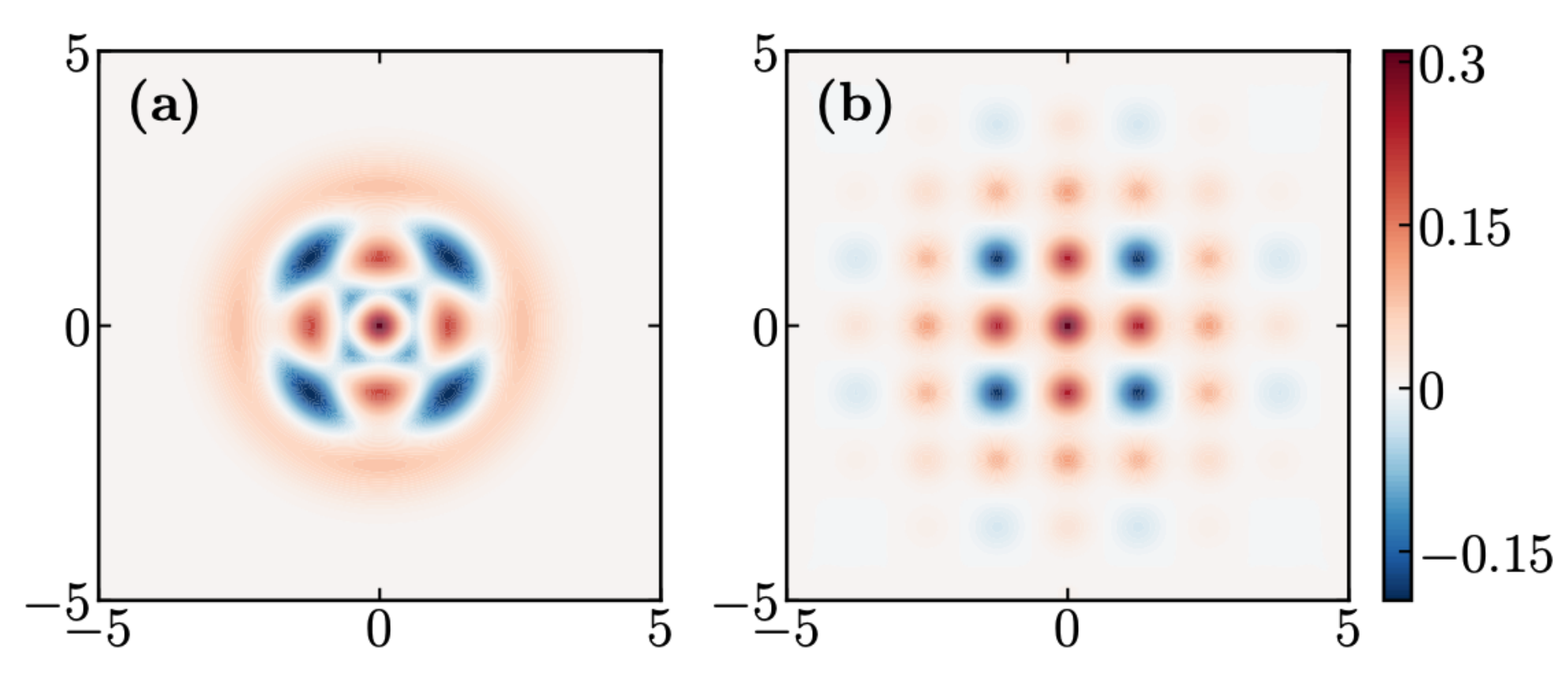}  

\caption{(a) Wigner function of the binomial state, defined in \eqref{eq:binomial-state}, that we use as input state. (b) Wigner function of the target qunaught state, with \unit[4.95]{dB} effective squeezing.}
\label{fig:wig_input_target}
\end{figure}

In particular, for the parameter values $K=3$ and $N=2$, we have
\begin{equation}
\label{eq:binomial-state}
\ket{\psi_0} \equiv \ket{0_2} = \frac{1}{2}\ket{0} + \frac{\sqrt{3}}{2}\ket{4},
\end{equation}
where we have introduced the notation $\ket{\psi_0}$ because this will be the input state of our conversion protocol. 
This state encodes the qubit state $0$. Note that what is referred to as binomial state encoding $0$-type logical information differs in different works, e.g., in Refs.~\cite{grimsmo_quantum_2020,nehra_all-optical_2021} it corresponds to $1/2\ket{0} + \sqrt{3}/2\ket{1}$ and in Ref.~\cite{axline2018demand} to $1/\sqrt{2}(\ket{0} + \ket{4})$. We will adopt the definition in \eqref{eq:binomial-state} throughout the rest of this paper. 
The Wigner function of this binomial state is plotted in \figpanel{fig:wig_input_target}{a}.

%---------------------------------------------
%---------------------------------------------

%--------------------------------------------------------------

\subsection{Qunaught states}
\label{sec:qunaught states}

The qunaught state, also called grid state, canonical GKP state, or sensor state, as it allows for detecting small displacements in phase space~\cite{duivenvoorden2017single}, is defined as~\cite{gottesman_encoding_2001}
\begin{equation}
%\ket{\psi} \propto \sum_{t=-\infty}^{\infty} e^{-\pi\Delta^2 t^2} D(t\xi/\sqrt{2}) S(\Delta) \ket{0} ,
\ket{\psi} \propto \sum_{t=-\infty}^{\infty} e^{-\pi\Delta^2 t^2} \hat D(t \sqrt{\pi}) \hat S(\Delta) \ket{0} ,
\end{equation}
where %$S(\Delta)=\textrm{exp}(\frac{1}{2}(\Delta^*\hat{a}^2-\Delta\hat{a}^\dagger^2))$ 
$\hat S(\Delta)$ 
is the squeezing operator which yields $\hat{q}$ $\rightarrow \hat{q}\Delta$ and $\hat{p}$ $\rightarrow \hat{p}/\Delta$, %$S(z)=S(\textrm{log}(\Delta))=\textrm{exp}(\frac{1}{2}(z^*\hat{a}^2-z\hat{a}^\dagger^2))$ 
and $\hat D(\beta)=\textrm{exp}(\beta \hat{a}^{\dagger}-\beta^*\hat{a})$ is the displacement operator, with $\hat a$ ($\hat a^\dag$) the annihilation (creation) operator of the bosonic mode. With this definition, the qunaught state has a spacing of $\sqrt{2\pi}$ between its peaks. As already mentioned, the qunaught state does not contain quantum information, as it is associated with a one-dimensional subspace.
 %The number of dimensions of the subspace depends on the choice of $\xi$ of the symmetric stabilizers $S_p=e^{i\xi \hat{p}}$ and $S_q=e^{i\xi \hat{q}}$~\cite{gottesman_encoding_2001}
%-----------------------------------------------------------------------------

One measure of the quality of a grid state with density matrix $\hat \rho$ is the effective squeezing~\cite{weigand_generating_2018}
\begin{equation}
\delta=\frac{1}{\sqrt{\pi}}\sqrt{\textrm{ln} \mleft| \textrm{Tr} [\hat D(\sqrt{\pi}) \hat\rho] \mright|^{-2}},
\label{eq:eff}
\end{equation}
where for an actual grid state one has $\delta = \Delta$.
In this paper, we will consider a target qunaught state with  squeezing $\Delta = 0.4$, corresponding to $S_{\textrm{GKP}}= \unit[4.95]{dB}$, 
where  $S_{\textrm{GKP}}= - 10 \textrm{log}_{10} (\Delta^2/\Delta^2_0)$~\cite{noh_low_2022}, with $\Delta^2_0=1/2$ the variance of the quadrature noise for the vacuum. %The effective squeezing of our target state is $0.4$.
 The Wigner function of this qunaught state is visualised in \figpanel{fig:wig_input_target}{b}.

%-----------------------------------------------------------------------------
%%%%%% Final check by Anton up to here %%%%%%
\section{QP breeding protocol}
\label{QP-breeding protocol}

Our iterative protocol for converting input binomial states into the target qunaught state is inspired by the breeding protocol of Ref.~\cite{weigand_generating_2018}. However, instead of starting from input squeezed-cat states, we start from the binomial state in \eqref{eq:binomial-state} and we consider alternating $\hat{q}$ and $\hat{p}$ measurements. 

\begin{figure}
\includegraphics[width=0.8\linewidth]{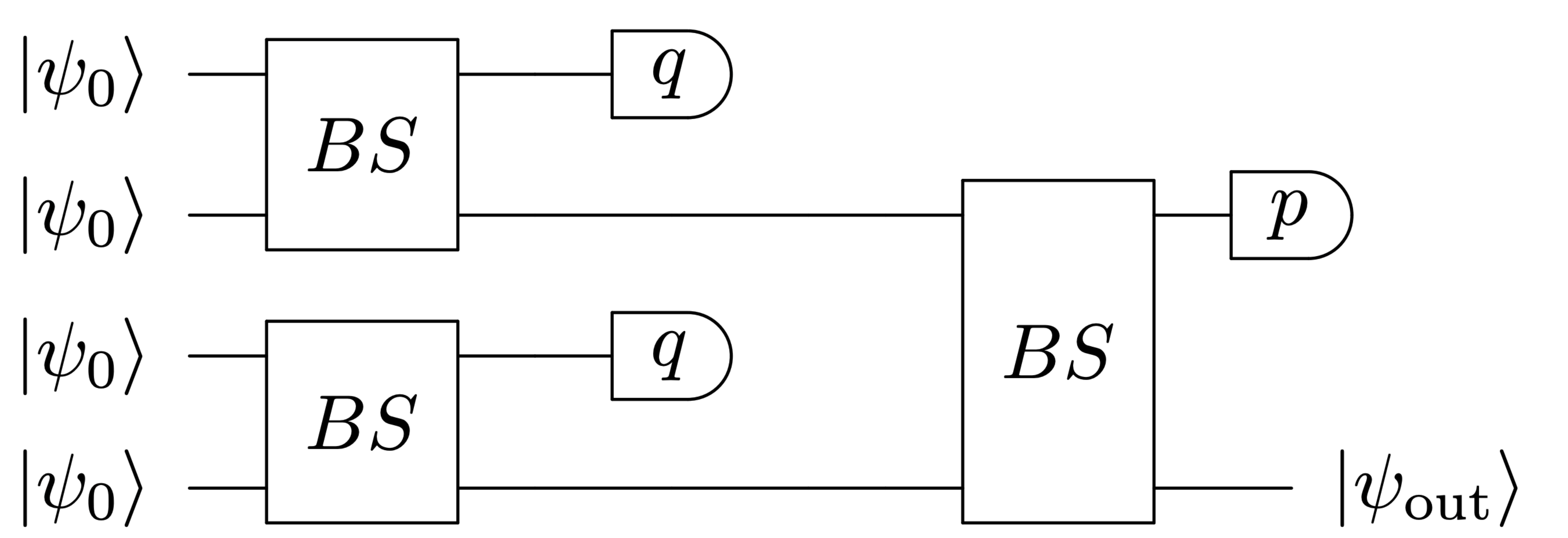}  
\caption{Sketch of our Gaussian conversion protocol with two iterations.
Two binomial states are combined in a real balanced (50:50) beam-splitter. We then use a homodyne detector to measure the first modes' $\hat{q}$ quadrature. The second mode yields the output state, which is used as the input state of the second iteration. In the second iteration, the measurement is done in the $p$ quadrature to induce the symmetry of the grid states.}
\label{fig:circuit2}
\end{figure}

Figure~\ref{fig:circuit2} shows the circuit implementing two iterations of our protocol. In the first iteration, two pairs of input binomial states $\ket{\psi_0}$ are entangled in real balanced beam-splitters. The first modes' position quadratures $\hat{q}$ are then measured. The resulting state in the second mode at the output of the each beam-splitter becomes the input state of the next iteration. As we investigate further in \appref{app:q-and-p-effect}, each $\hat{q}$ and $\hat{p}$ measurement induces a squeezing on the output state in $\hat{q}$ and $\hat{p}$ quadratures, respectively. Therefore, in the second iteration, we measure the $\hat{p}$ quadrature to balance the squeezing strength in the two directions.

\begin{figure}
\includegraphics[width=1\linewidth]{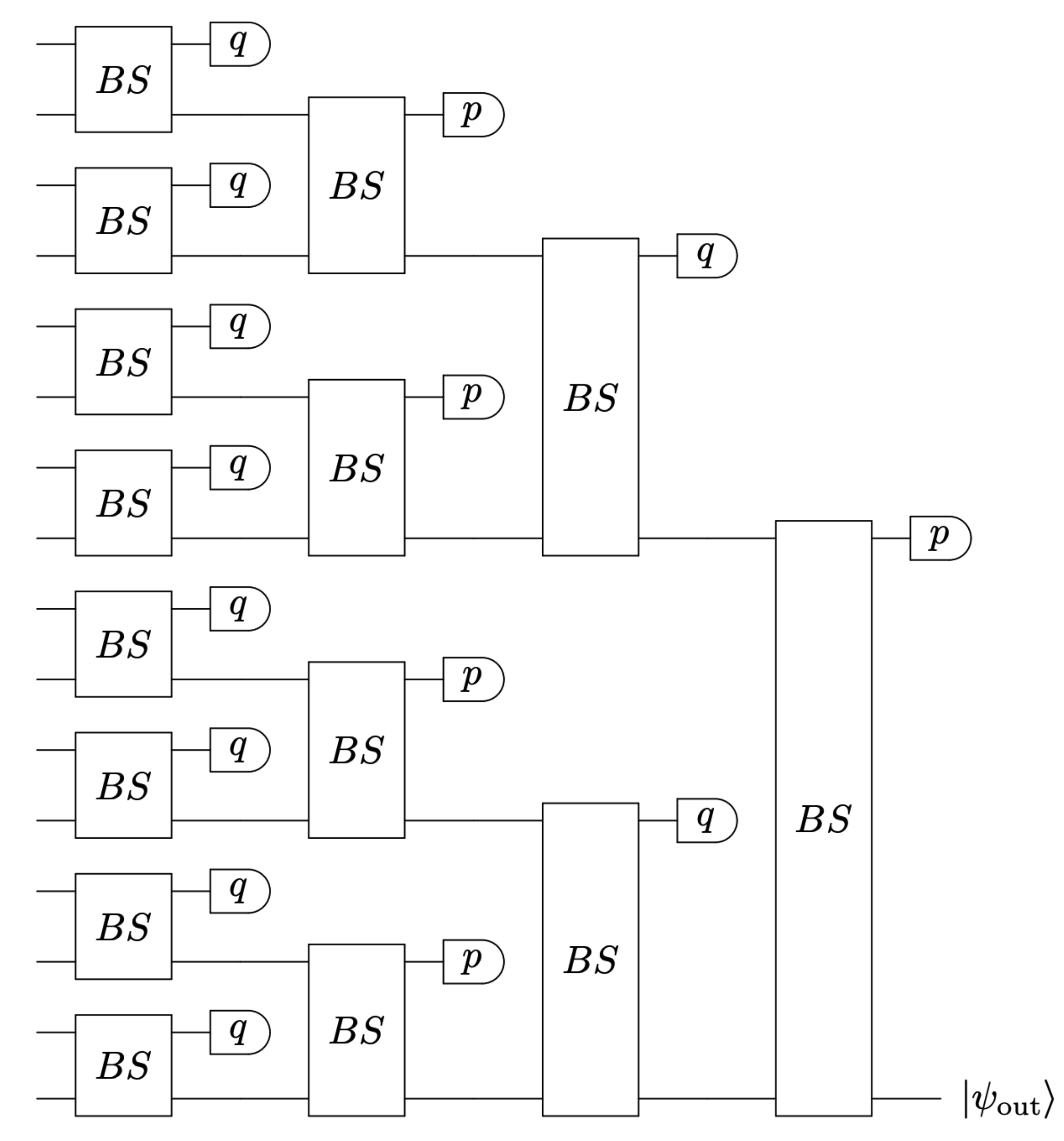}  
\caption{Sketch of our Gaussian conversion protocol using multiple iterations of the building-block circuit in \figref{fig:circuit2}.
}
\label{fig:circuit_many}
\end{figure}

Figure~\ref{fig:circuit_many} illustrates how to generalize the unit circuit of \figref{fig:circuit2} to four iterations of the protocol. 
%If we also do the momentum measurement in each iteration, we will be approaching a squeezed state after 2 iterations.
The resulting output state $\ket{\psi_{\rm out}}$ depends on the measurement outcomes obtained at each measurement. Extending the protocol to even more iterations is done following the same pattern from the unit circuit in \figref{fig:circuit2}.

%---------------------------------------------
%---------------------------------------------

%---------------------------------------------
%---------------------------------------------

In what follows, we will be interested in two different scenarios. In the first scenario, we consider post-selection on values of the homodyne result that are close to zero all the time. In the second scenario, we allow for post-selection on different values. As we will see, generally speaking, the first protocol will yield higher fidelity to the target qunaught state, and lower success probability, than the second protocol.

In order to simulate the measurement in the position and momentum basis, we construct corresponding observables within a finite Hilbert space. We truncate the Hilbert space dimension to 50, consider the matrix representation of the position and momentum quadratures in the Fock basis, and diagonalise them numerically to obtain the eigenvectors and eigenvalues. Since the number operator is phase invariant, the eigenvalues are the same for $\hat{q}$ and $\hat{p}$. In particular, the eigenvalues closest to zero are $q/\sqrt{2 \pi} = \pm 0.062$ and  $p/\sqrt{2 \pi} = \pm 0.062$. The obtained eigenvectors are used to implement the projective quadrature measurements considered in this paper.

%---------------------------------------------------------------------------

\section{Performance analysis}
\label{sec:result}

In this section, we numerically evaluate the performance of our QP-breeding protocol, focusing first on the output state obtained after the first iteration, then on the output state obtained after two iterations, and finally on the output states after multiple iterations.

%--------------------------------------------------------------

\subsection{First iteration}

\begin{figure}
\includegraphics[width=1\linewidth]{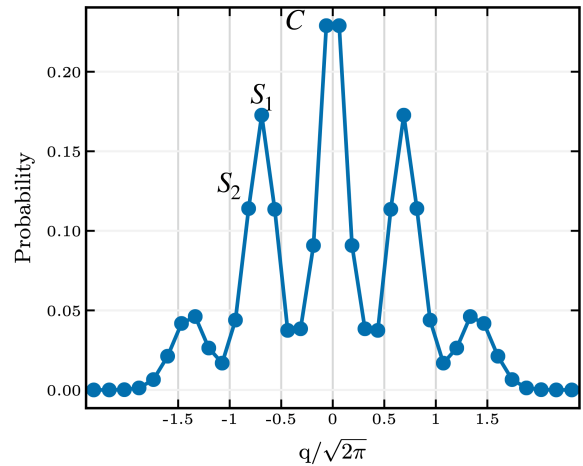}  
%$C_1:q=-0.156=0.0622\sqrt{2 \pi}$, $S_1:q=-1.728=0.0689\sqrt{2 \pi}$, $S_2=-2.046=0.816\sqrt{2 \pi}$ 
\caption{Probability of measuring the rescaled outcome $q/\sqrt{2 \pi}$ at the first homodyne measurement in the circuit of \figref{fig:circuit2} for input binomial states as defined in \eqref{eq:binomial-state}. The values $q/\sqrt{2 \pi}=-0.062$ (denoted $C$, since it is the central peak), $q/\sqrt{2 \pi}=-0.689$ (denoted $S_1$, since it is a side peak), and $q/\sqrt{2 \pi}=-0.816$ (denoted $S_2$, since it is a further point close to the side peak) are the most probable outcomes. 
}
\label{fig:iter1prob}
\end{figure}

In \figref{fig:iter1prob}, we show the probability distribution of the measurement outcomes of the first $\hat{q}$ measurement. We can see that the most probable measurement outcomes correspond to a central peak and two side peaks. More specifically, the three highest probabilities are at the central peak $C$, at the top of a side peak $S_1$, at the point next to the top of a side peak $S_2$, and at their respective mirror images with respect to the axis $q=0$~. Together, these measurement outcomes occur with a probability of \unit[51.56]{\%} \footnote{The peak symmetric to $S_2$ with respect to $S_1$ corresponds to an outcome equally probable as $S_2$, but yields lower fidelities to the target qunaught state, in either one or more rounds of the protocol, see \appref{sec: all cases}.  Therefore we will not consider this value further.}.  

\begin{figure}
\centering
\includegraphics[width=1\linewidth]{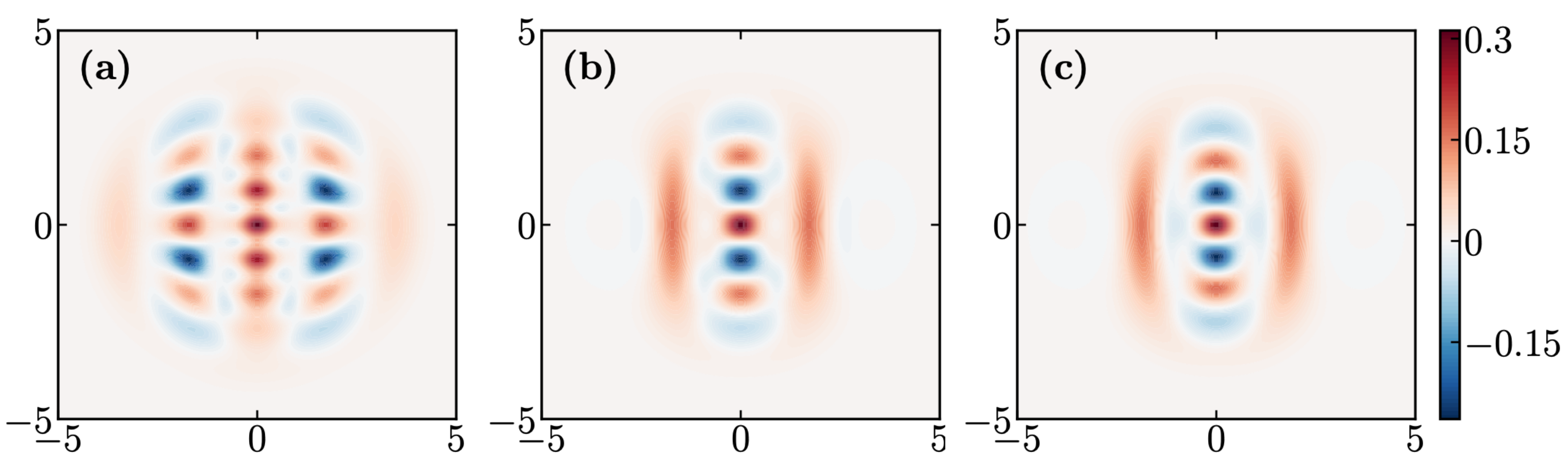}
\caption{Wigner functions of the output states after the first iteration, corresponding to the peak points (a) $C$, (b) $S_1$, and (c) $S_2$, of the probability distribution shown in \figref{fig:iter1prob}.}
\label{fig:wig_1iter}
\end{figure}

The Wigner functions of the state in the unmeasured output mode for these three cases are shown in \figref{fig:wig_1iter}. 
In \figpanel{fig:wig_1iter}{a}, we see that the state obtained for a measurement outcome corresponding to the central peak of the distribution displays a grid structure. However, the fidelity of this state with a qunaught state is only \unit[73.0]{\%} (with a corresponding success probability of \unit[22.9]{\%}). To further increase the fidelity we need to increase the number of iterations.
In \figpanels{fig:wig_1iter}{b}{c}, we show the states obtained for measurement outcomes corresponding to the side peaks $S_1$ and $S_2$, respectively. These states resemble squeezed cat states. Note that squeezed cat states are the input states in the protocol of Ref.~\cite{weigand_generating_2018}, where $\hat{p}$ measurements are used iteratively. This hints that we will approach a qunaught state in the next iteration with a measurement in $\hat{p}$ quadrature.

%---------------------------------------------

%---------------------------------------------
%---------------------------------------------

\subsection{Two iterations}

Next, we study the output state after two iterations of the protocol, i.e., after one round of homodyne detection in $\hat{q}$ and one in $\hat{p}$ have been performed, as illustrated in \figref{fig:circuit2}. Since there are 50 possible outcomes for each quadrature measurement with our truncation of the Fock space, there are $50^3 =$ 125,000 possible sequences of measurement outcomes when using two iterations. 
However, as we saw in Fig.~\ref{fig:iter1prob}, the most probable measurement outcomes (the peaks) in the first iteration are $C$, $S_1$, $S_2$, and their symmetric points with respect to the axis $q = 0$. Furthermore, the symmetry of the probability distribution along the axis $q=0$ allows us to only consider half of these cases (either positive or negative $q$).

\begin{figure}
\includegraphics[width=1\linewidth]{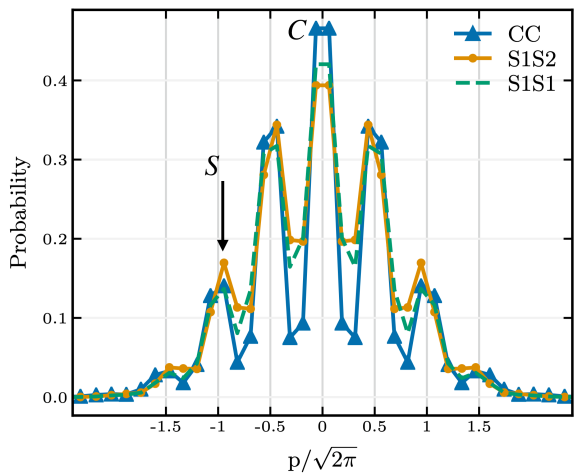}  
\caption{Probability of measuring the rescaled outcome  $p/\sqrt{2 \pi}$ in the second homodyne measurement in the circuit of \figref{fig:circuit2}, for the most probable outcomes of the first homodyne measurement (in $\hat{q}$ quadrature), corresponding to the peaks in \figref{fig:iter1prob}. The case $S_1S_2$ is identical to the case $S_2S_1$. The peak $S$ corresponds to the value $p/\sqrt{2 \pi}=-0.944$. }
%Ps: They all look quite the same in terms of the location of the peaks. However, only the cases of CC and S1S2 have the fidelity above 98\%. The others' fidelities are in the range [95\%-97\%]}
\label{fig:iter2prob}
\end{figure}

Moreover, the probability distribution of the $\hat{p}$-measurement outcomes in the second iteration displays a similar structure as for the $\hat{q}$ measurement at iteration level one, with well-defined peaks. This is illustrated in \figref{fig:iter2prob} for the most probable outcomes obtained in the first iteration. Here we introduce a new label $S$ to indicate a new side peak in the $\hat{p}$-measurement probability distribution, as shown in \figref{fig:iter2prob}~\footnote{We will disregard the peak point located between $C$ and $S$ as the output state obtained by post-selecting on the corresponding measurement value does not yield high fidelity.}.
%We will use the same labels $S_1$, $S_2$ and $C$ to indicate the four most probable outcomes (along with their symmetric points with respect to the $q$ axis). 
We can thus focus our analysis on the cases that occur with higher probability, and analyze the corresponding output states in the unmeasured mode.

%---------------------------------------------
%---------------------------------------------

\begin{figure}
\centering
\includegraphics[width=1\linewidth]{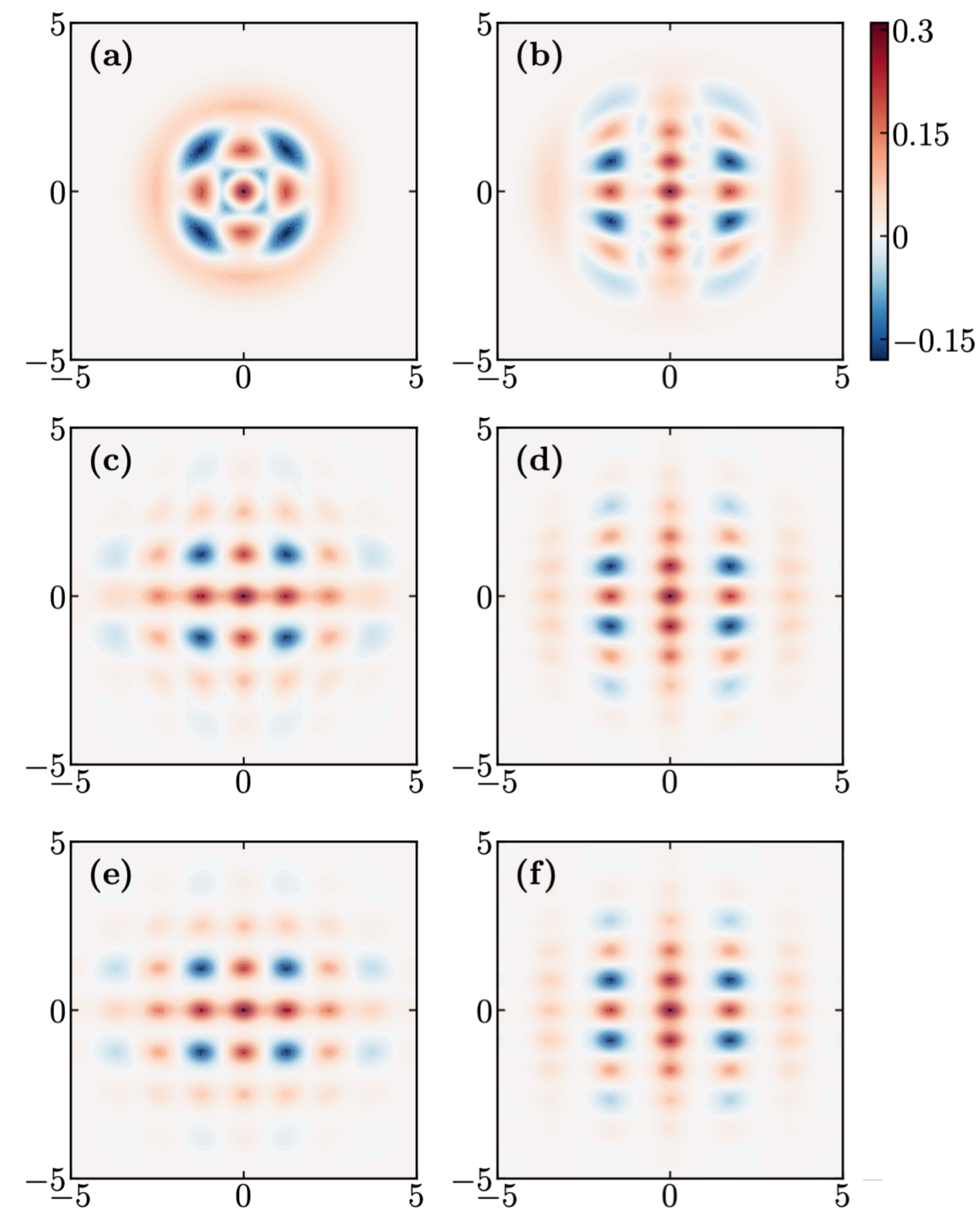}

\caption{Wigner functions of (a) the input state [defined in \eqref{eq:binomial-state}] and (b-f) the output states generated by the first five iterations, respectively, of the QP protocol, when all the measurement values %are $\pm 0.15$
are post-selected to be $0.062 \sqrt{2 \pi}$. The target state is fixed to be the \unit[4.95]{dB}-squeezing qunaught state shown in \figpanel{fig:wig_input_target}{b}.}
\label{fig:wigner_multi_0}
\end{figure}

\begin{figure}
\includegraphics[width=\linewidth]{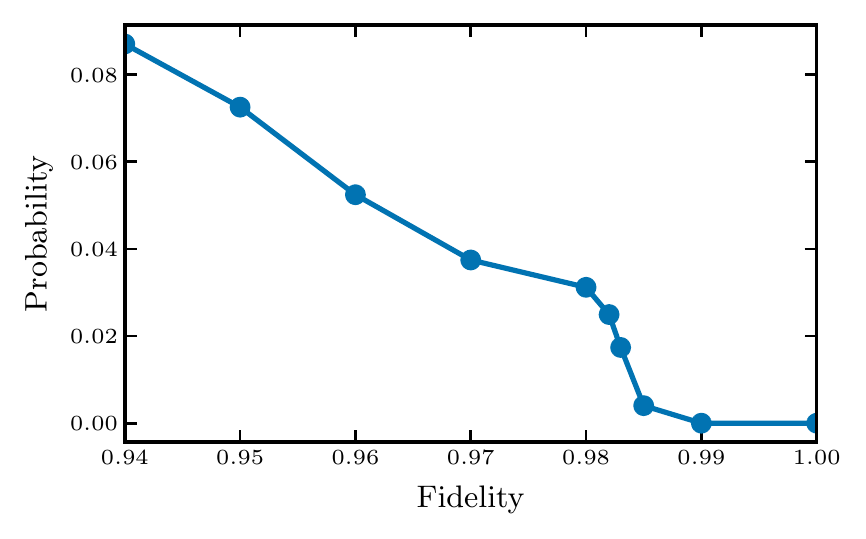}  
\caption{Success probability as a function of the lower bound on the fidelity between the protocol's output states and the target qunaught state. Here, the output states are generated after two iterations.}
\label{fig:twoiter_fid}
\end{figure}

In Table~\ref{table:2iter}, we show the fidelity and probability of occurrence, along with the effective squeezing, for  some of the most probable cases after two iterations. As we can see, in the case labelled by $CCC$, where we post-select on the central peak for each of the measurements, we can reach a fidelity of \unit[98.34]{\%}, and this case occurs with \unit[1.3]{\%} probability. The Wigner function of the corresponding output state is shown in \figpanel{fig:wigner_multi_0}{c}. A slightly higher fidelity, but a lower success probability, is achieved in the $CCS$ case. All in all, two iterations of the QP protocol allow for achieving a fidelity above \unit[98]{\%} with a probability of \unit[3.14]{\%}, where we post-select on $q$ and $p$ values located at the center or side peaks considered. In \figref{fig:twoiter_fid}, we show the success probability with which various values of the fidelity to the target state can be reached after two iterations; the highest fidelity with non-zero success probability is \unit[98.87]{\%}, the two right-most points in \figref{fig:twoiter_fid} corresponding instead to a zero success probability within the numerical precision of our calculation. A more extensive table with cases, corresponding fidelities, and probabilities of achieving the target state with a fidelity above 96\% is provided in \appref{sec: all cases}.
%---------------------------------------------
% make it look like pra table: Hline.

\begin{table}[h]
\begin{center} {%\footnotesize
\begin{tabular}{cccc}
\hline
\hline
\\[-1ex]
 {Peaks} & {Fidelity} & {Probability} & {Effective squeezing}\\[0.5ex]
 \hline 
 \\[-0.5ex]
 $C$ $C$ $S$ & 0.9887 & 0.004 & 0.3765 \\
 $C$ $C$ $C$ & 0.9834 & 0.0134 & 0.3522 \\ 
 $S_1$ $S_2$ $C$ & 0.9830 & 0.0038 & 0.4918 \\
 $S_1$ $S_1$ $C$ & 0.9816 & 0.0063 & 0.5003 \\[.3ex]
\hline
\end{tabular} }
\end{center}
\caption{ Fidelity to the \unit[4.95]{dB}-squeezing qunaught state, probability, and effective squeezing for the output state generated after two iterations of the QP protocol with a few different measurement results. For all these cases, the fidelity is above 0.98. For comparison, the effective squeezing obtained from Fig.~5 of Ref.~\cite{weigand_generating_2018} is between 0.3 and 0.4.}
\label{table:2iter}
\end{table}

%---------------------------------------------

%---------------------------------------------
%---------------------------------------------

\begin{figure}
\includegraphics[width=\linewidth]{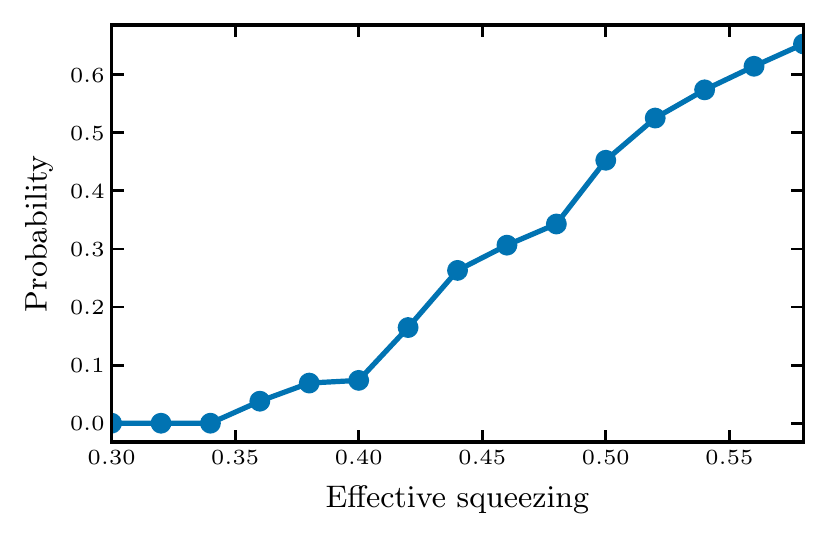}  
\caption{Success probability as a function of the upper bound of the effective squeezing after two iterations.}
\label{fig:twoiter_eff_sq}
\end{figure}

As we introduced in Eq.~(\ref{eq:eff}), the effective squeezing $\delta$ is another parameter that reflects how well a state is approaching our target state. Figure~\ref{fig:twoiter_eff_sq} shows the probability to achieve an output state with effective squeezing below a certain value with two iterations of our protocol. We observe that there is above \unit[7.4]{\%} (\unit[30]{\%}) probability to obtain an output state with effective squeezing below 0.4 (0.46). For reference, the input binomial state's effective squeezing is 0.53, while, as we mentioned, the target qunaught state's effective squeezing is 0.4. We thus have a non-zero probability to reach an even stronger effective squeezing than in the target state, which is linked to the appearance of more peaks in the wave-function of the output state than those in the wave-function of the target state, as shown in Appendix~\ref{app:best}. 

%---------------------------------------------
%--------------------------------------------------------------

\subsection{More iterations}

Considering qunaught states as simultaneous eigenstates of two commuting displacement operators, it is natural to study the case of further iterations of projection measurements in the $\hat{q}$ and $\hat{p}$ quadratures. In \figref{fig:wigner_multi_0}, we show the Wigner function of the output state that results when post-selecting on the lowest eigenvalue $0.062 \sqrt{2 \pi}$ in all measurements for one to five iterations. By comparing with \figpanel{fig:wig_input_target}{b}, we see that the output states gradually approach a qunaught state after each even number of iterations, i.e., when the symmetry in the two directions in phase space is preserved.

%---------------------------------------------

%---------------------------------------------

In Table~\ref{table:many_iter_0}, we show the fidelity and the success probability after even numbers of iterations, when we post-select on the lowest eigenvalue at every measurement. We see that while the fidelity can reach values above 0.989, the corresponding success probability decreases rapidly for large numbers of iterations.

%--------------------------------------------
%---------------------------------------------

\begin{table}[h]
\begin{center} {%\footnotesize
\begin{tabular}{ccc}
\hline
\hline
\\[-1ex]
 {Iteration} & {Fidelity} & {Probability} \\[0.5ex]
 \hline 
 \\[-3ex]\\
 0 & 0.941 & 1 \\
2 & 0.9848 &  0.0134\\ 
4 & 0.9875 & $4.5\times 10^{-10} $ \\
6 & 0.9884 & $6.4\times 10^{-40} $ \\
8 & 0.9897 & $6.4\times 10^{-159}$  
\\[.3ex]
\hline
\end{tabular} }
\end{center}
\caption{ Fidelity to the 4.95 dB-squeezing qunaught state and probability for the output state generated after
0 to 8 iterations, when post-selecting on the lowest eigenvalue at each $\hat{q}$ and $\hat{p}$ measurement.}
\label{table:many_iter_0}
\end{table}

In Appendix~\ref{app:Numerical-N-K}, we investigate how the output fidelity changes when we consider as input other binomial states, i.e., identified by other rotational symmetry order $N$ or truncation parameter $K$.

%---------------------------------------------------------------------------

\section{Conclusion}
\label{sec:conclusion}
We have developed a protocol for the heralded generation of qunaught states, which is inspired by the scheme described in Ref.~~\cite{weigand_generating_2018}. However, our protocol uses input binomial states and alternating quadrature measurements in combination with beamsplitters, rather than relying on squeezed cat states and on fixed $\hat{p}$-measurements as in the original scheme. Our protocol produces states that are approximately eigenstates of the commuting displacement operators, resulting in qunaught states. Using numerical simulations, we have demonstrated that it is possible to achieve qunaught states with a fidelity of above \unit[98]{\%} with a probability of \unit[3.14]{\%} after only two iterations of the protocol. While it is possible to achieve higher fidelities with more iterations of the protocol, this comes at the cost of lower success probabilities.

Our scheme shows that binomial states can be Gaussian-converted to grid states (an hence in turn, with further probabilistic Gaussian operations, to a computational basis $\ket{0}$ GKP state), and as such they are universal in combination with Gaussian resources. Previously, they were only known to be universal in combination with other non-Gaussian resources, such as further auxiliary specific encoded binomial states and non-linear operations~\cite{grimsmo_quantum_2020}. More in general, binomal $\ket{0}$ states therefore add to the relatively short list of non-Gaussian states that can potentially provide all-Gaussian and fault-tolerant universality. Specifically, in Ref.~\cite{baragiola2019all} it has been demonstrated that ideal zero-logical encoded GKP qubits promote circuits composed of Gaussian elements to universality and fault tolerance via magic-state (Gaussian) distillation. The robustness of such a result to small deviations would therefore entail that also squeezed cat states could provide all-Gaussian fault-tolerant universality, via Gaussian conversion to zero-logical encoded GKP qubits as envisaged in the protocols introduced in Refs.~\cite{vasconcelos_all-optical_2010, etesse_proposal_2014, weigand_generating_2018}. The same would be true for single-photon states --- via Gaussian conversion to squeezed cat states~\cite{etesse2015experimental} --- and for binomial states --- via the conversion protocol proposed in this work. One interesting question that stems from our work is which other non-Gaussian states promote the Gaussian tool-box of operations to universal quantum computation. The study of conversion protocols can shed light on this question. 

%---------------------------------------------------------------------------

\section{Acknowledgements}

We thank Oliver Hahn, Shahnawaz Ahmed, and Timo Hillmann for fruitful discussions.
YZ, AFK, and GF acknowledge support from the Knut and Alice Wallenberg Foundation through the Wallenberg Centre for Quantum Technology (WACQT). GF acknowledges support from the VR (Swedish Research Council) Grant QuACVA. 

%-----------------------------------------------------------------------------
\appendix

%%%%%%%%%%%%%%%%%%%%%%%%%%%%%%%% Checked by Anton up to here %%%%%%%%%%%%%%%%%%%%%%%%%%%%%%%%

\section{Effect of $\hat q$ and $\hat p$ measurements in the QP breeding protocol}
\label{app:q-and-p-effect}

\begin{figure}
\centering

\hfill
\includegraphics[width=1\linewidth]{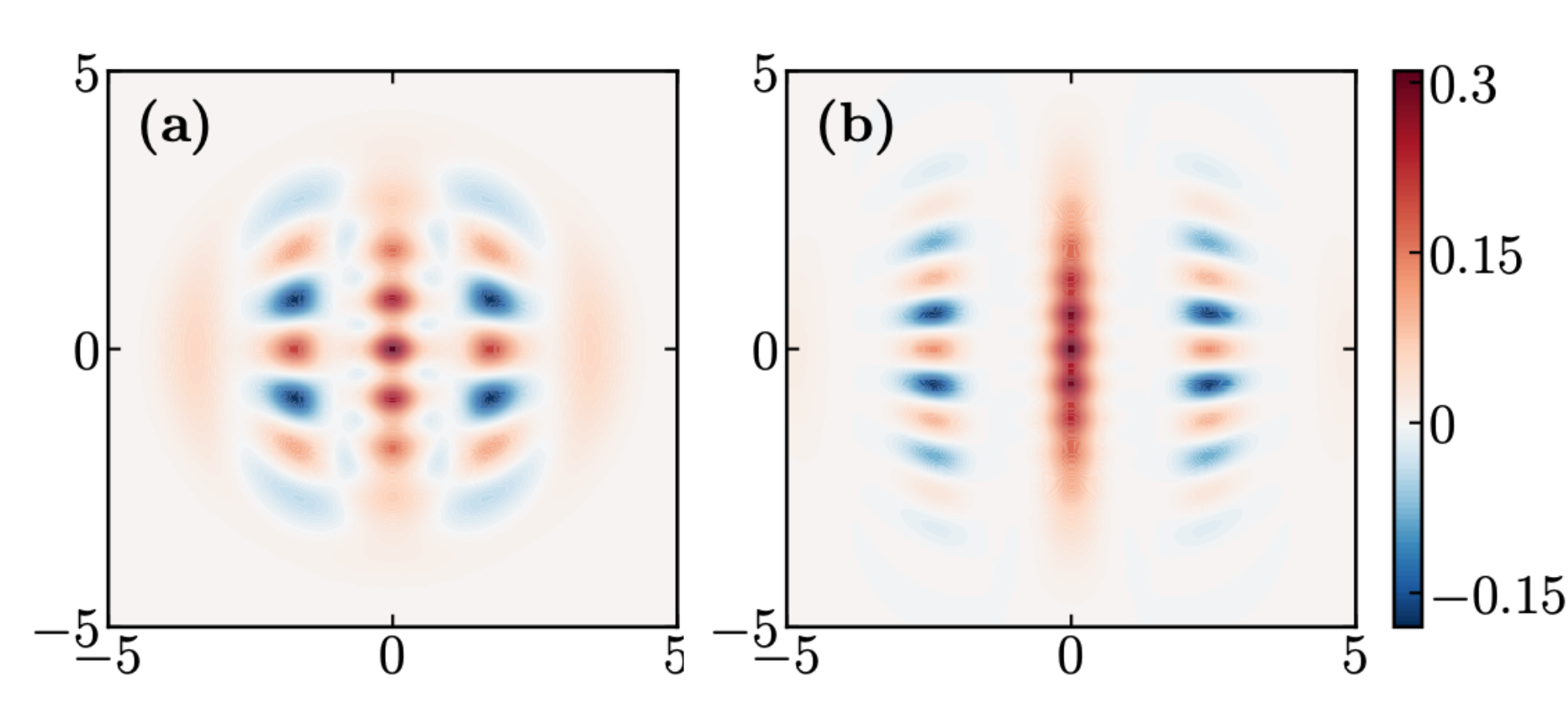}
\caption{Wigner function of the output state after (a) one and (b) two iterations of the protocol in \figref{fig:circuit2}, but with the measured modes all being measured along the quadrature $\hat{q}$.}
\label{fig:q_wig_meas}
\end{figure}

\begin{figure}
\centering
\includegraphics[width=1\linewidth]{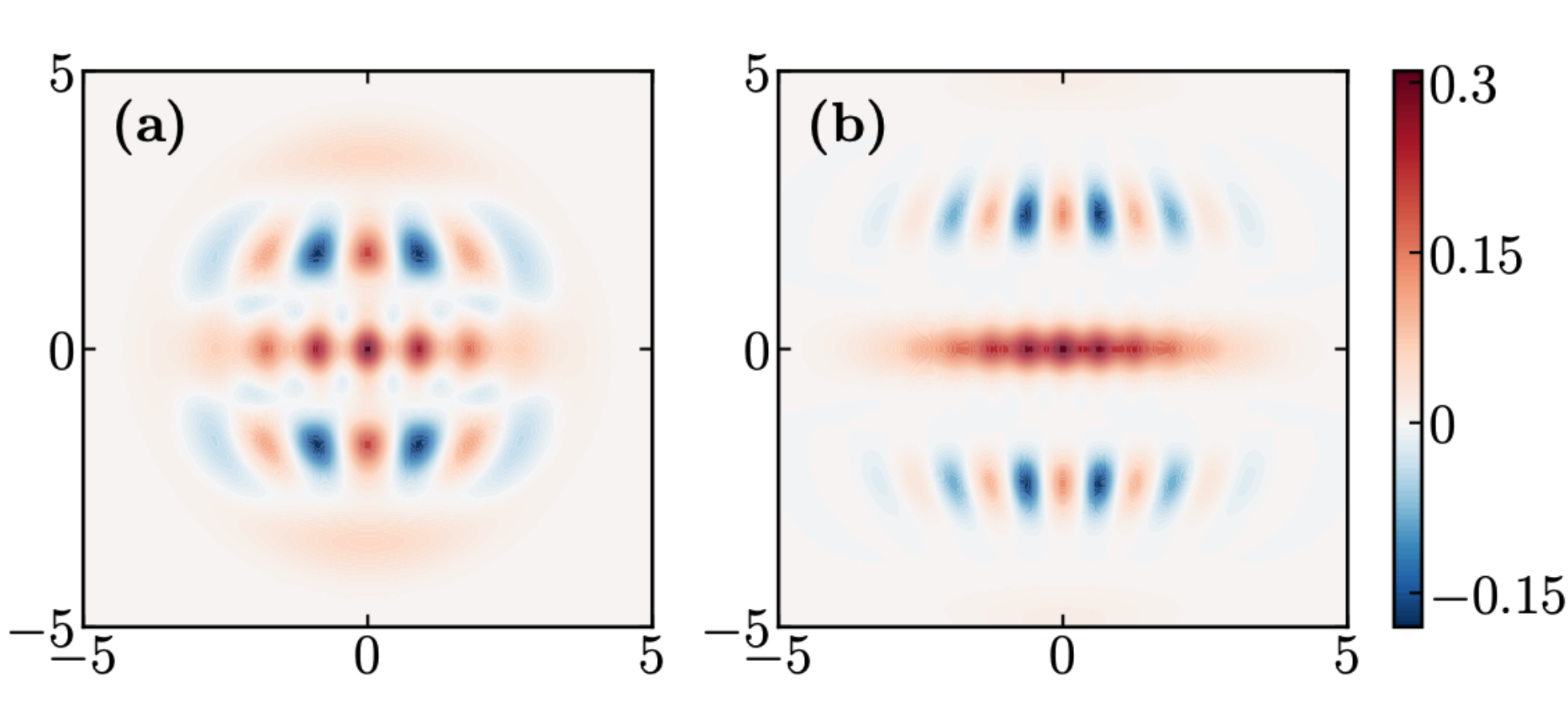}
\hfill
\caption{Wigner function of the output state after (a) one and (b) two iterations of the protocol in \figref{fig:circuit2}, but with the measured modes all being measured along the quadrature $\hat{p}$.}
\label{fig:p_wig_meas}
\end{figure}

Figures~\ref{fig:q_wig_meas} and \ref{fig:p_wig_meas} show the effect of sequential $\hat{q}$ and $\hat{p}$ measurements, in case the same quadratures are measured repeatedly (instead of in an alternating way). We see that each $\hat{q}$ and $\hat{p}$ measurement induces a squeezing on the output state in $\hat{q}$ and $\hat{p}$ quadratures, respectively, and this guides us in the choice of alternating $\hat{q}$ and $\hat{p}$ measurements.

%---------------------------------------------
%---------------------------------------------

%---------------------------------------------
%---------------------------------------------

%---------------------------------------------------------------------------

\section{Fidelity to target qunaught states depending on the measurement outcomes after two iterations}
\label{sec: all cases}

\begin{figure}
\centering
\includegraphics[width=\linewidth]{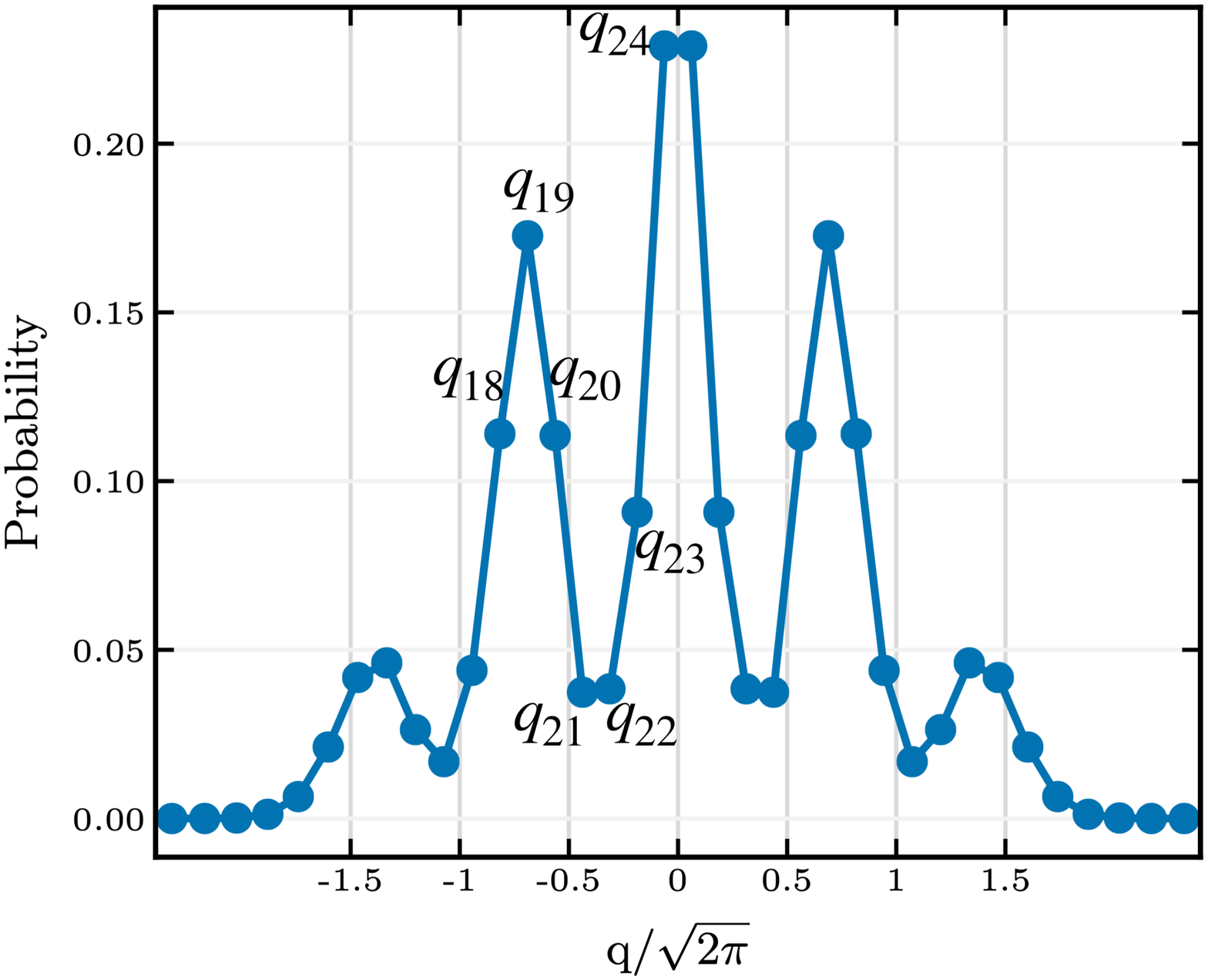}
\hfill
\includegraphics[width=\linewidth]{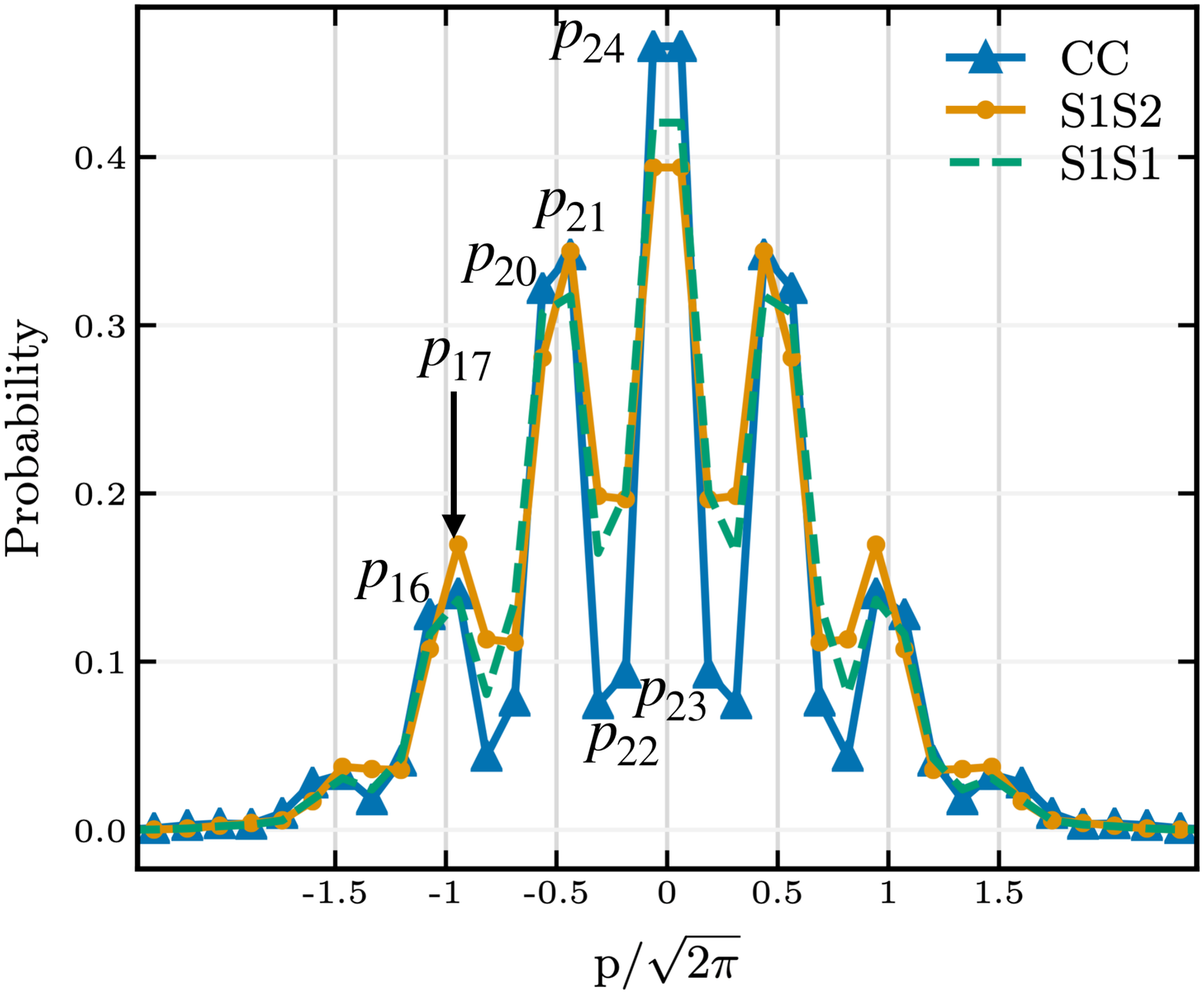}
\caption{Top (bottom) panel: Probability of measuring the rescaled outcome $q/\sqrt{2 \pi}$ ($p/\sqrt{2 \pi}$) in the first (second) homodyne measurement in the circuit of \figref{fig:circuit2} for input binomial states. We label the peaks using the label of the corresponding Fock eigenvalue, for the first 25 ($i \in [0,24]$) of the 50 eigenvalues.}
\label{fig:p_wig_meas_relabel}
\end{figure}

In Table~\ref{table:2iter_all}, we show the fidelity and probability of occurrence, along with the effective squeezing, for all combinations of $\hat{q}$ and $\hat{p}$ measurement outcomes yielding a fidelity above \unit[96]{\%} to the target qunaught state after two iterations of the QP protocol. Since we here consider more cases than in Table~\ref{table:2iter}, we introduce a new notation for indicating the peaks. This notation is defined in the two panels of
%Fig~\ref{fig:p_wig_meas_relabe
\figref{fig:p_wig_meas_relabel}.

%---------------------------------------------
\begin{table}
\begin{tabular}{cccc}
\hline
\hline
\\[-1ex]
{Peaks} & {Fidelity}  & {Probability} & {Effective squeezing}\\[1ex] 
\hline
\\[-1ex] 
%\hline
$q_{18}q_{18}p_{16}$ & 0.963 & 0.0006 & 0.4806 \\
$q_{18}q_{19}p_{24}$ & 0.983 & 0.0038 & 0.4918 \\
$q_{18}q_{20}p_{24}$ & 0.9723 & 0.0021 & 0.4828 \\
$q_{19}q_{18}p_{24}$ & 0.983 & 0.0038 & 0.4918 \\
$q_{19}q_{19}p_{16}$ & 0.968 & 0.0017 & 0.5022 \\
$q_{19}q_{19}p_{17}$ & 0.9721 & 0.002 & 0.4894 \\
$q_{19}q_{19}p_{24}$ & 0.9816 & 0.0063 & 0.5003 \\
$q_{19}q_{20}p_{24}$ & 0.9659 & 0.0036 & 0.5133 \\
$q_{20}q_{18}p_{24}$ & 0.9723 & 0.0021 & 0.4828 \\
$q_{20}q_{19}p_{24}$ & 0.9659 & 0.0036 & 0.5133 \\
$q_{20}q_{20}p_{16}$ & 0.9661 & 0.0011 & 0.5061 \\
$q_{20}q_{20}p_{17}$ & 0.9644 & 0.0008 & 0.4832 \\
$q_{24}q_{24}p_{16}$ & 0.9669 & 0.0037 & 0.4185 \\
$q_{24}q_{24}p_{17}$ & 0.9887 & 0.004 & 0.3765 \\
$q_{24}q_{24}p_{24}$ & 0.9834 & 0.0134 & 0.3522  \\
[1ex] 
\hline
\end{tabular}
\caption{Fidelity to the \unit[4.95]{dB}-squeezing qunaught state, probability, and effective squeezing for the output state generated after two iterations of the QP protocol with the measurement results that yield fidelities above 0.96.}
\label{table:2iter_all}
\end{table} 
%---------------------------------------------

%---------------------------------------------

\section{Probability distribution in the position representation of the best output state after two iterations}
\label{app:best}

\begin{figure}
\centering
\includegraphics[width=1\linewidth]{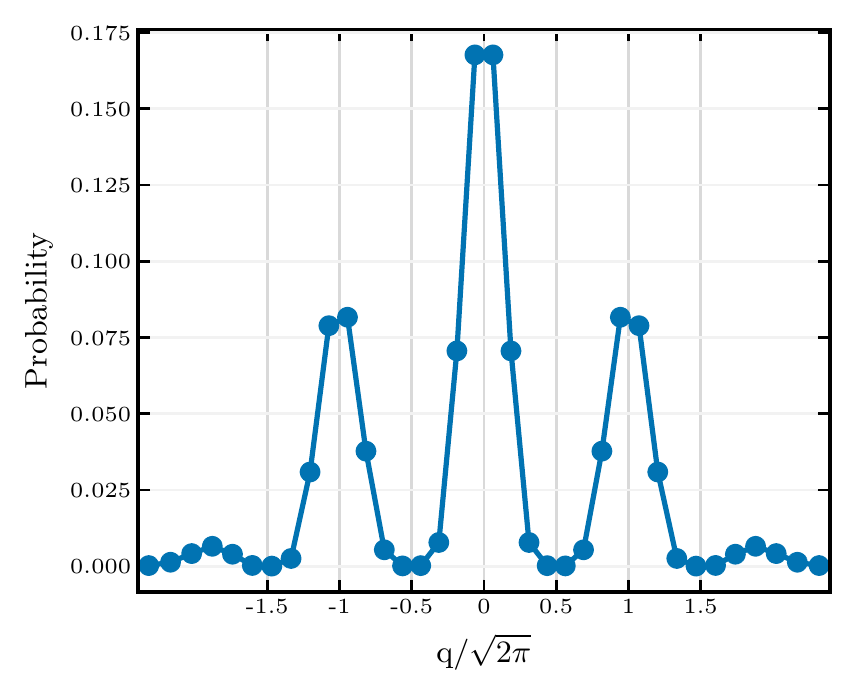}
\caption{Probability distribution in the position representation of the best output state after two iterations of the QP protocol.}
\label{fig:prob242417}
\end{figure}

As shown in Table~\ref{table:2iter}, the highest fidelity is achieved in the $CCS$ case within two iterations of the protocol.
In Fig.~\ref{fig:prob242417}, we plot the probability distribution in the position representation of the corresponding best output state.
%---------------------------------------------

%---------------------------------------------------------------------------

\section{Fidelity to target qunaught states depending on order of rotational symmetry and truncation parameter of the input binomial state}
\label{app:Numerical-N-K}

\begin{figure}
\includegraphics[width=1\linewidth]{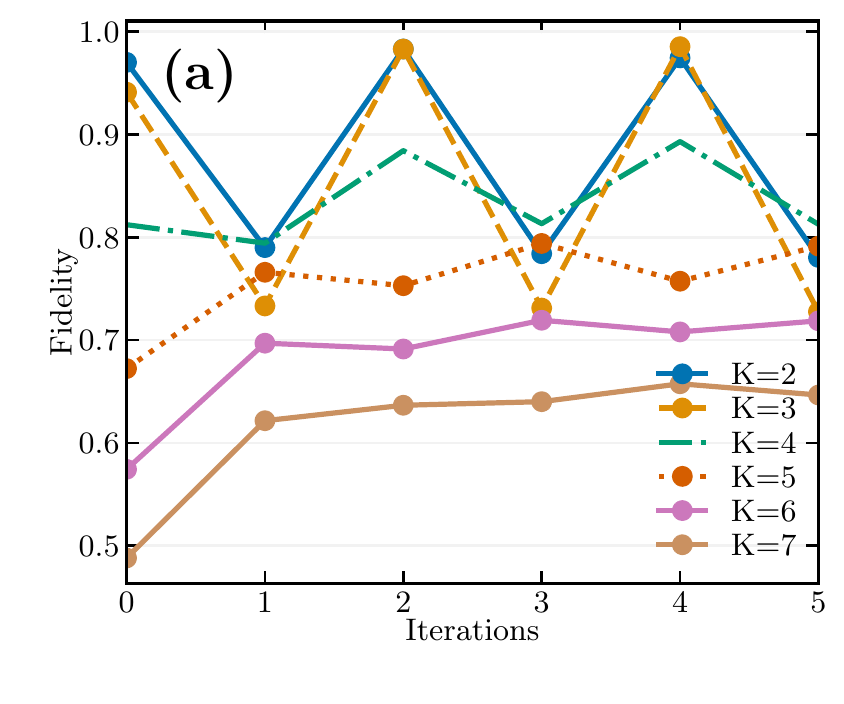}   
\includegraphics[width=1\linewidth]{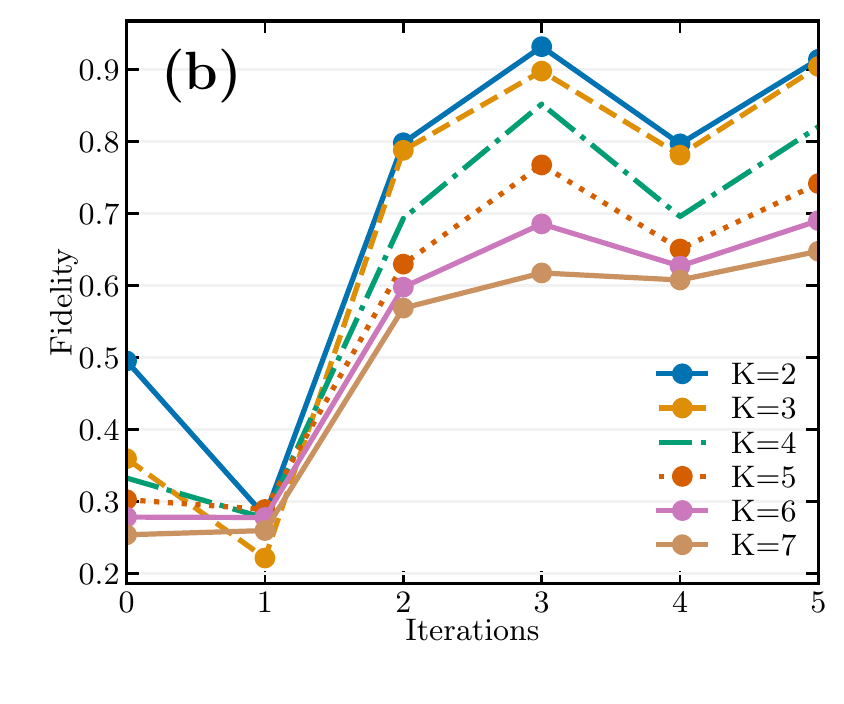} 
\includegraphics[width=1\linewidth]{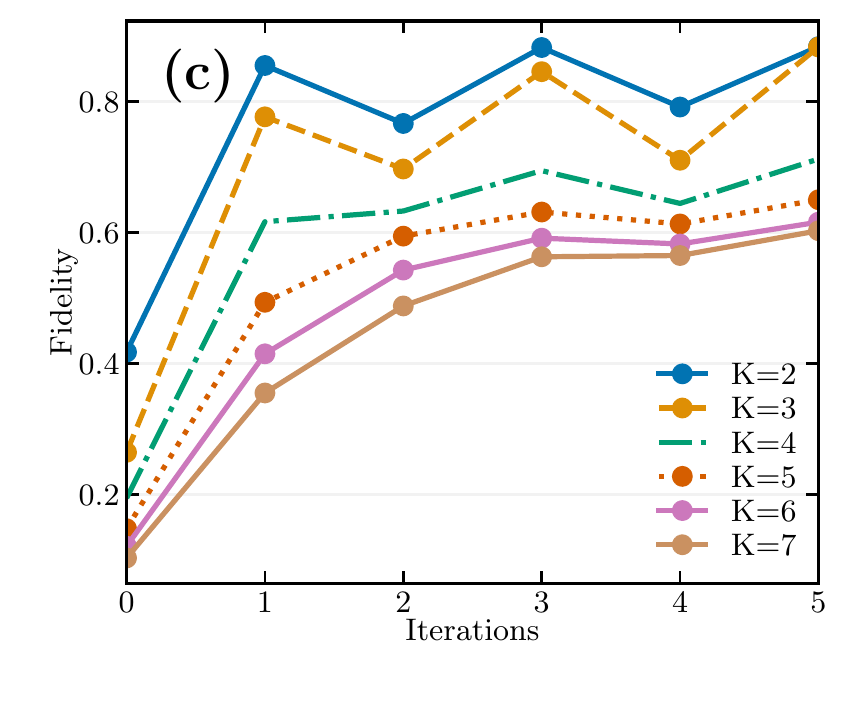} 
\caption{Fidelity to a target qunaught state with squeezing $\Delta = 0.4$ as a function of the number of iterations for binomial input states with rotational symmetry of order (a) $N=2$, (b) $N=3$, and (c) $N=4$, and for $K = 2 - 7$ for each $N$.}
\label{fig:gadget}
\end{figure}

Here we consider other binomial states than that of \eqref{eq:binomial-state} as input for the QP protocol. In \figref{fig:gadget}, we plot the fidelity to the same target qunaught state as in the main text as a function of the number of iterations of the QP protocol, considering both even and odd numbers of iterations of the $\hat{q}$ and $\hat{p}$ measurements, for other input binomial states. These other binomial states correspond to various values of the order of the rotational symmetry $N$, as well as of the truncation parameter $K$. The oscillating behaviour of the fidelity to the target qunaught state as a function of the iteration number corroborates our choice of a set of alternating $\hat{q}$ and $\hat{p}$ measurements as the basic building block for our iterative protocol. This figure also informs us that the highest fidelities are achieved for the case that we studied in the main text, i.e., $N = 2$ and $K = 3$, which is hence a sweet spot for targeting the generation of qunaught states with this protocol. From the figure, we also note that a comparable result in terms of the fidelity is obtained in the case of $K = 2$.
% $\ket{\psi_0} =\frac{1}{2}\ket{0}+\frac{\sqrt{3}}{2}\ket{4}$.

\begin{figure}
\includegraphics[width=1\linewidth]{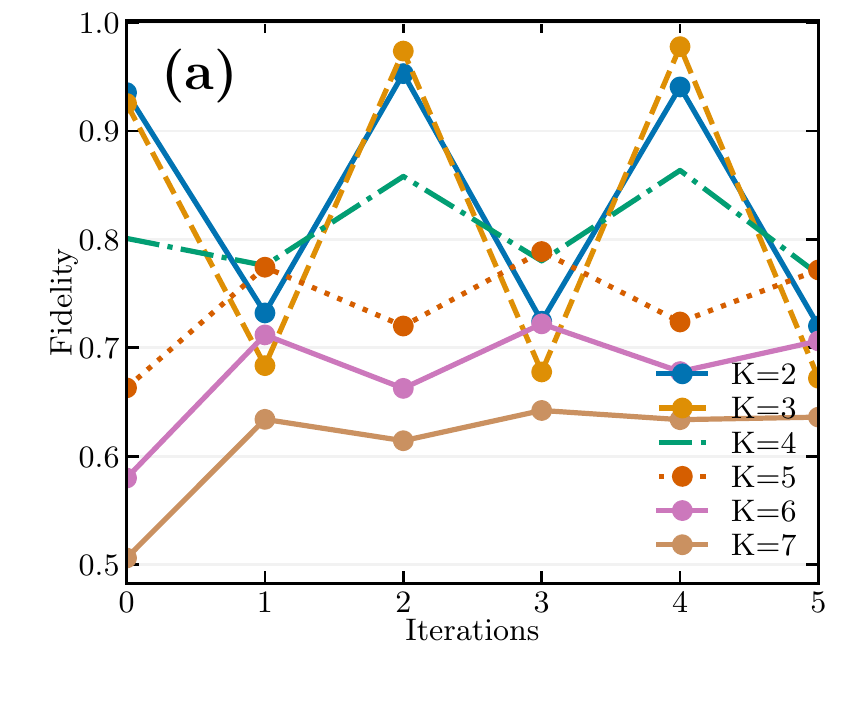}   
\includegraphics[width=1\linewidth]{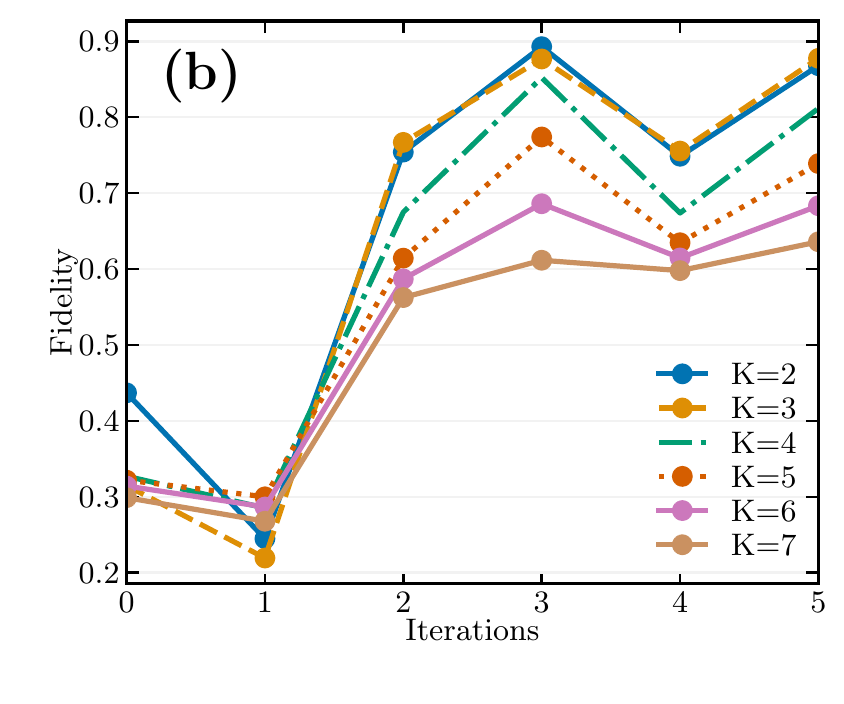} 
\includegraphics[width=1\linewidth]{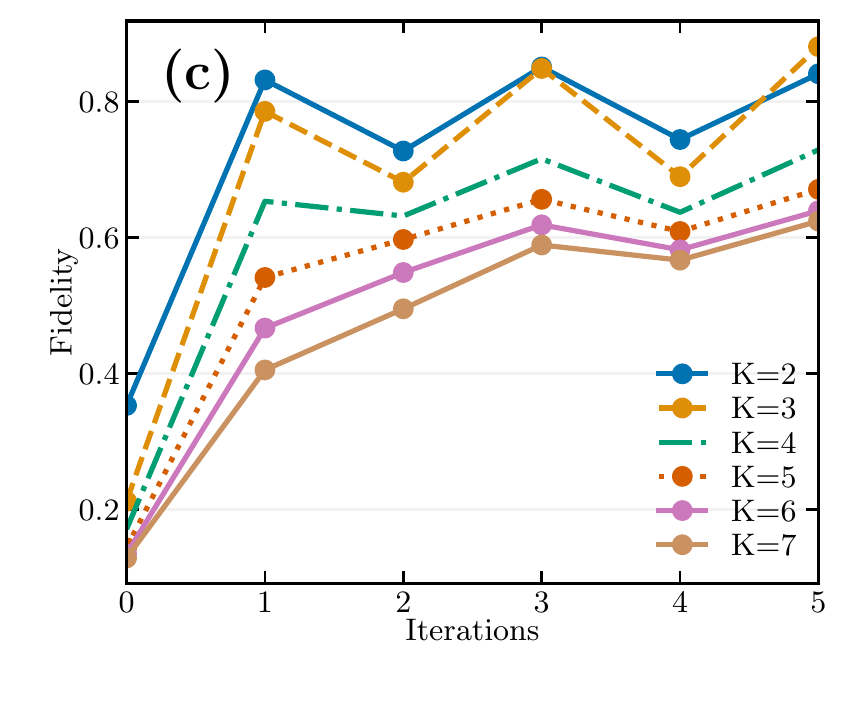} 
\caption{Fidelity to a target qunaught state with squeezing $\Delta = 0.35$ as a function of the number of iterations for binomial input states with rotational symmetry of order (a) $N=2$, (b) $N=3$, and (c) $N=4$, and for $K = 2 - 7$ for each $N$.}
\label{fig:gadget035}
\end{figure}

All these observations also appear to be valid if we choose a slightly different target qunaught state. In \figref{fig:gadget035}, we illustrate this robustness by plotting the same quantities as in \figref{fig:gadget} for a target qunaught state with squeezing $\Delta = 0.35$. Our intuition for this result is that the crucial factor determining the best input state is the spacing between peaks in the input state, which affects the spacing between peaks in the output state. Changing the squeezing of the target qunaught state does not change the spacing between peaks in that state.

%---------------------------------------------

%---------------------------------------------
\bibliographystyle{apsrev4-1}
\bibliography{bib}

\end{document}